\title{Observations of Anomalous Microwave Emission from HII regions}
\author{
        Clive Dickinson\\
        Jodrell Bank Centre for Astrophysics\\
        School of Physics \& Astronomy\\
        University of Manchester\\
        Oxford Road, Manchester, M13 9PL (U.K.)
}

\date{\today}

\documentclass[letterpaper,12pt, ]{article}
\usepackage{epsfig}
\begin{document}

\maketitle

\begin{abstract}
In this brief review, I give a summary of the observations of Anomalous Microwave Emission (AME) from HII regions. AME has been detected in, or in the vicinity of, HII regions. Given the difficulties in measuring accurate SEDs over a wide range of frequencies and in complex environments, many of these detections require more data to confirm them as emitting significant AME. The contribution from optically thick free-free emission from UCHII regions may be also be significant in some cases. The AME emissivity, defined as the ratio of the AME brightness to the $100\,\mu$m brightness, is comparable to the value observed in high-latitude diffuse cirrus in some regions, but is significantly lower in others. However, this value is dependent on the dust temperature. More data, both at high frequencies ($>\sim 5$\,GHz) and high resolution ($\sim 1^{\prime}$ or better) is required to disentangle the emission processes in such complex regions. 
\end{abstract}

\section{Introduction}
\label{sec:intro}
HII regions refer to the environment around the most (O and B type) massive stars, which are hot enough to produce intense UV radiation that can ionize the gas around them. HII regions typically form within large molecular clouds, often in clusters (due to triggered star formation), and are therefore are also associated with significant amounts of dust grains. Anomalous Microwave Emission (AME), if due to electric dipole radiation from spinning dust \cite{Draine1998a}, requires a large column of dust grains (with a population of the smallest dust grains or PAHs) and a mechanism for rotationally exciting these grains e.g. plasma drag, photons etc. For these reasons, HII regions may be a good place to look for AME. In fact, there is evidence that photodissociation regions (PDRs) typically found around the edges of HII regions/molecular clouds might be good AME emitters \cite{Casassus2008,PEP_XXI}. Counter-arguments include the depletion of PAHs close in the centre of HII regions and the fact that they strongly emit in other forms of continuum emission, notably free-free (thermal bremmsstrahlung) and thermal dust radiation.

In this brief review, I give an overview of the the continuum radiation and current observations of AME from HII regions. I will discuss some issues with measuring AME from HII regions, including calibration, the contribution from ultracompact (UCHII) regions and the definition of emissivity.

\section{Observations of HII regions}

\subsection{The SEDs of classical HII regions}
\label{sec:classic_hii}

\begin{figure*}
    \centering
    \includegraphics[width=9cm, angle=0]{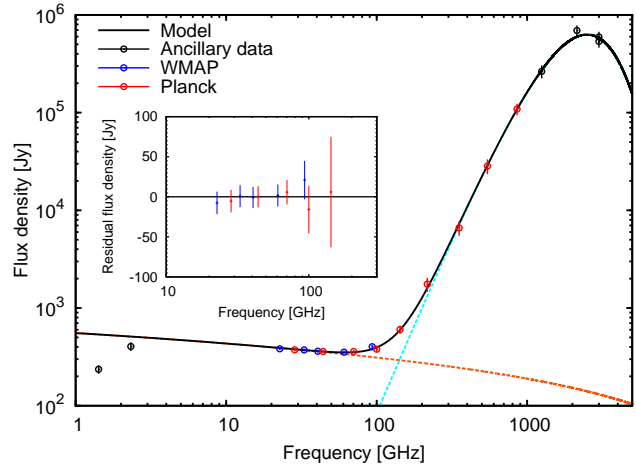}
\caption{SED of the Orion Nebula (M42) HII region \cite{PEP_XX}. Free-free emission dominates at frequencies below 100\,GHz while thermal dust emission dominates above 100\,GHz. The free-free emission is optically thin above a few GHz. There is no evidence of significant AME.}
\label{fig:m42}
\end{figure*}

The general form of HII region SEDs (radio to the far infrared continuum) is thought to be well understood. Fig.~\ref{fig:m42} shows the SED of the well-known Orion nebula (M42) HII region, measured by a number of different experiments including {\it Planck} \cite{PEP_XX}. The spectral shape is typical of HII regions. At radio wavelengths (frequencies $\sim 1$\,GHz to $\sim 100$\,GHz) it is dominated by free-free emission from warm ($T_e\sim 10^4$\,K) ionized gas. This is usually expressed in terms of the free-free opacity, $\tau_{ff}$, which at radio wavelengths can be approximated by \cite{Dickinson2003}:

\begin{equation}
\tau_{ff} \approx 3.27 \times 10^{-7} \left( \frac{T_e}{10^4\,{\rm K}} \right)^{-1.35} \left( \frac{\nu}{{\rm GHz}} \right)^{-2.1} \left (\frac{EM}{{\rm pc\,cm}^{-6}} \right) .
\end{equation}

The intensity is proportional to the Emission Measure, defined as $EM=\int n_e^2dl$, the integral of the square of the electron density along the line of sight. Above a certain ``turn-over'' frequency (i.e. $\tau_{ff}<1$; optically thin), free-free emission has an almost flat flux density spectrum of $\alpha \approx -0.1$ ($S \propto \nu^{\alpha}$). At lower frequencies, and particularly for more dense (young and compact HII regions) with $EM>>10^7$\,pc\,cm$^{-6}$, the emission becomes optically thick ($\tau_{ff}>1$) and has a spectrum of $\alpha=+2$. 

At frequencies $\ge \sim 100$\,GHz, black-body emission from dust grains at $T_d \sim 10$--$100$\,K dominates. The thermal dust spectrum is often parameterised by a modified black-body function, $S \propto \nu^{\beta +2}B(\nu,T_d)$, with typical values of the emissivity index of $\beta \sim +1.8$ in the Rayleigh-Jeans tail (corresponding to $\alpha=+3.8$) and a peak at $\sim 3000$\,GHz ($100\,\mu$m). Also, the dust temperatures around HII regions are typically warmer ($\sim 30$--80\,K) compared to the diffuse cirrus ($T_d \sim 18$\,K) \cite{Dupac2003}.

\subsection{Observations of AME in HII regions}
\label{sec:ame_obs}

To search for AME, one is essentially looking for excess emission at frequencies $\sim 30$\,GHz. AME is usually detected at frequencies in the range $\sim 10$--60\,GHz where the non-AME components are weaker. Also, theoretical models of spinning dust tend to peak at frequencies near 30\,GHz \cite{Draine1998a}. Table~\ref{tab:obs} summarises the observations of AME from HII regions to-date\footnote{We focus on observations that have either detected AME or have placed upper limits on AME. We do not include results from reflection nebulae or PNe. There exists many other observations at frequencies relevant to AME in the literature, most of which do not show obvious signature of AME (most of these are at relatively high angular resolution with instruments such as the GBT, VLA and ATCA).}. We list the frequency range, approximate angular scales (of the experiment or of the source, whichever is largest), the AME significance level ($\sigma_{\rm AME}$) and the emissivity ($E$), defined as the AME brightness relative the $100\,\mu$m brightness, in units of $\mu$K/(MJy/sr). The list is ordered in terms of their approximate AME detection significance level.

\bigskip
\begin{table*}
\footnotesize
\caption{Summary of observations of AME from HII regions. The entries are listed according to their approximate detection significance, $\sigma_{\rm AME}$. The angular sizes are approximate, or refer to the telescope beam. $E$ is the AME emissivity relative to $100\,\mu$m, in units $\mu$K/(MJy/sr). $^{a}$Emissivity for W40 esimate is based on a $100\,\mu$m flux density of $10^5$\,Jy and a $2\,\sigma$ upper limit at 33\,GHz of 5\,Jy.}
\label{tab:obs}
\begin{tabular}{lccccccl}
\hline \\
Source(s)           &Experiment   &$\nu$    &$\theta$   &$\sigma_{\rm AME}$ &$E$                      &Ref.(s)                         &Notes   \\
                    &             &[GHz]    &[$^{\prime}$] &                  &                         &                                & \\
\hline \\
G159.6--18.5       &Various      &10--60   &60         &17                &$17.8^{\pm0.3}$            &\cite{Watson2005,PEP_XX}          &Perseus MC. Low free-free  \\
RCW175              &CBI/VSA      &31       &10         &7.9               &$5.5^{\pm0.7}$             &\cite{Dickinson2009,Tibbs2012}    &2 components \\
G173.6+2.8        &{\it Planck} &30--70   &60         &6.4               &$10.0^{\pm3.4}$            &\cite{PEP_XX}                     &Contains S235 \\
9 northern HII      &VSA          &33       &$\sim 15$         &5.4               &$3.9^{\pm0.8}$             &\cite{Todorovic2010}              & Mean \\
G107.1+5.2        &{\it Planck} &30--70   &60         &4.8               &$11.3^{\pm4.8}$            &\cite{PEP_XX}                     &Contains S140 \\
RCW49               &CBI          &31       &8          &3.3               & $13.6^{\pm4.2}$           &\cite{Dickinson2007}              & G284.3--0.3 \\
6 southern HII      &CBI          &31       &$\sim 5$--10      &1.9               &$3.3^{\pm1.7}$             &\cite{Dickinson2007}              & Mean includes RCW49 \\
W40                 &WMAP/CBI     &31--33   &60/8         &5$^*$                 &$<0.2^{a}$                          &\cite{Finkbeiner2004}             &$^*$No excess detected by \cite{Stamadianos2010} \\
LPH96~201.6+1.6     &GBT/CBI      &5--30    &$\sim 10$         &10$^*$               &$5.8^{\pm2.3}$             &\cite{Finkbeiner2002,Dickinson2006} &$^*$Spurious; AME $< 24$\,\% \cite{Dickinson2006} \\
16 compact HII      &AMI          &15       &$<1$       &$\sim 0$       &$<5$                      &\cite{Scaife2008}                 &Upper limits only \\
\hline \\
\end{tabular}
\end{table*}
\normalsize

\section{Discussion}
\label{sec:discussion}

\subsection{Reliability of detections}
\label{sec:reliability}

The measurement of AME in HII regions is clearly a difficult task. It must be remembered that measuring accurate flux densities over a wide range of frequencies, particularly for extended regions in the presence of complicated backgrounds (as is often the case for Galactic HII regions), is exceptionally difficult. For low angular resolution observations, the free-free and dust emission regions will be coincident and therefore may well be a small fraction of the total flux. To subtract free-free to say 1\% precision relies on having absolute flux scales that are good to this accuracy (most astronomical data are accurate to a few \% and many older data are good to $\sim 10$\,\% or worse!). Perhaps even more problematic is the comparison of data with a wide range of angular resolutions, and especially interferometric data compared to single-dish data, where the response to different angular scales can vary and is difficult to quantify (unless a detailed model of the source is available).

Given the above issues, one must be cautious given that the majority of the detections listed in Table~\ref{tab:obs} are not hugely significant (i.e. they are $\sim 5\sigma$ or below). One of the most clear detections comes from the source G159.6--18.5 within the Perseus molecular cloud. However, this is actually a very weak HII region relative to the dust-correlated emission from the larger surrounding area (the {\it Planck} AME source is actually located at G160.26--18.62). The free-free emission is therefore a very small fraction ($\approx 10\,\%$) of the total large-scale flux. The environment is therefore somewhat different to the other HII regions and perhaps should not be compared with the other HII regions.

From the other HII regions listed in Table ~\ref{tab:obs}, several of them are likely to be spurious detections. LPH96~201.663+1.643 was one of the first claimed detections of AME \cite{Finkbeiner2002} based on a rising spectrum from 5 to 10\,GHz. However, it was later shown that this result is likely to be spurious when no AME was observed at 31\,GHz with an upper limit of 24\,\% ($2\,\sigma$) \cite{Dickinson2006}. Indeed, private communication with Doug Finkbeiner revealed that follow-up observations of this source with the GBT did not confirm the spectral rise seen in early observations. Similarly, an analysis of the SED of W40 using WMAP 1-year data \cite{Finkbeiner2004} suggested that W40 may have a significant AME excess at 33\,GHz. However a re-analysis using WMAP 7-year data, combined with higher resolution CBI data, could not confirm any significant deviations from an optically thin free-free spectrum \cite{Stamadianos2010}. The detection of RCW49 \cite{Dickinson2007} could be contested based on the reliability and scarcity of low frequency (1--15\,GHz) data. If one were to remove the data point at 14.7\,GHz, which happens to be lower than the other data, then the significance of the detection at 31\,GHz is reduced to $\sim 2\sigma$. Finally, analyses involving averaging the results from a sample of HII regions (e.g. \cite{Dickinson2007,Todorovic2010}) can be misleading since systematics errors (e.g. calibration, background subtraction etc) can become dominant.

Clearly, more data at a range of frequencies and angular resolution are required to confirm and improve the accuracy for the quantification of AME.

\subsection{Compact vs extended regions and the contribution from UCHII}
\label{sec:compact_extended}

The SED of an evolved diffuse HII region ($EM << 10^6$\,pc\,cm$^{-6}$) will typically be optically thin above $\sim 1$\,GHz. However, very compact HII regions, with $EM>10^7$\,cm\,pc$^{-6}$ (ultracompact (UCHII) and hypercompact (HCHII)) can have turnover frequencies of $\sim 15$\,GHz and higher. These would be difficult to detect at lower frequencies. A nearby ionized region at $T_e\sim10\,000$\,K with angular size $\sim\!1^{\prime\prime}$ could have a maximum flux density of up to $\sim\!10$\,Jy at 30\,GHz although most are at $<1$\,Jy \cite{Wood1988}. It is therefore possible that AME (or a portion of it) from HII regions could be produced by UCHII regions that turnover at $\sim 15$--40\,GHz.

Estimating the contribution from UCHII is somewhat difficult. Methods include using high resolution radio data to extrapolate flux densities of point sources assuming a given $EM$ and angular size (e.g. Perrott et al., this issue) or the use of H$\alpha$ \cite{Dickinson2003} and/or Radio Recombination Line(s) data \cite{Alves2012}. Another way is to use an empirical relation between the ratio of $100\,\mu$m flux density, $S_{\rm 100\,\mu{\rm m}}$,  and 2\,cm (15\,GHz) radio flux density, $S_{\rm 2\,cm}$, from \cite{Kurtz1994} who measured $S_{100\,\mu{\rm m}}/S_{\rm 2\,cm}$ values between 1000 and 360000, with no UCHII regions below 1000; the median value was $\sim 3000$--5000. We apply this method for the Perseus, S140 and S235 AME regions to estimate the maximum UCHII flux density, $S_{\rm max}$, assuming $S_{100\,\mu{\rm m}}/S_{\rm 2\,cm}=1000$.

To identify UCHII candidates within the vicinity of these HII regions, we use the colour-colour relation of \cite{Wood1989} who found that UCHII regions tend to have IRAS colour ratios of ${\rm log}_{10}(S_{60}/S_{12}) \geq 1.30$ and ${\rm log}_{10}(S_{25}/S_{12})\geq 0.57$. Although this method was found to be very useful for finding the majority of UCHII regions, it also selects a large fraction of non-UCHII regions, such as cloud cores with lower mass stars \cite{Ramesh1997}. This therefore serves to be a very conservative upper limit to the contribution of UCHII, and is more likely to be a significant over-estimate by factors of several.

\begin{figure*}
    \centering
    \includegraphics[width=5.4cm, angle=0]{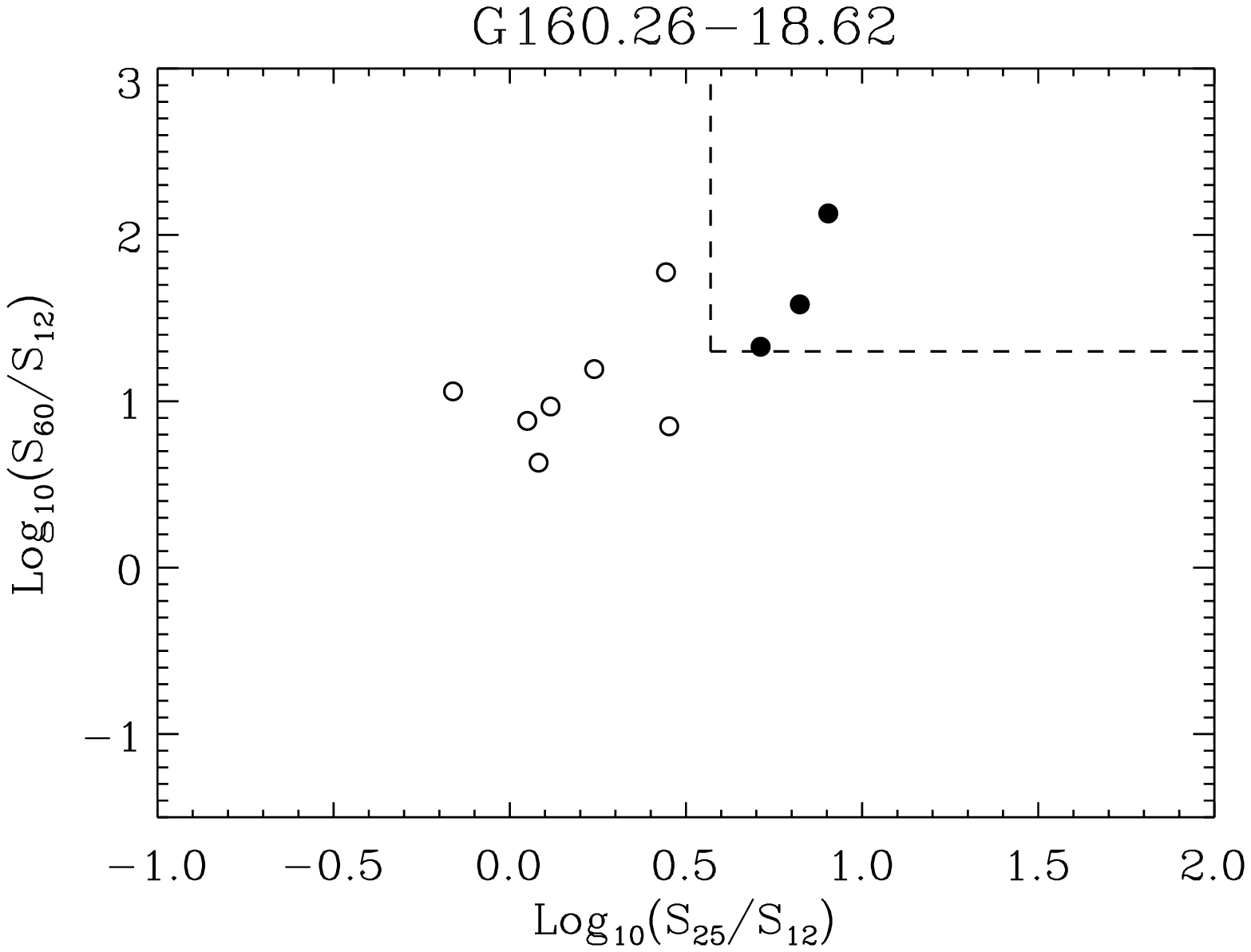}
    \includegraphics[width=5.4cm, angle=0]{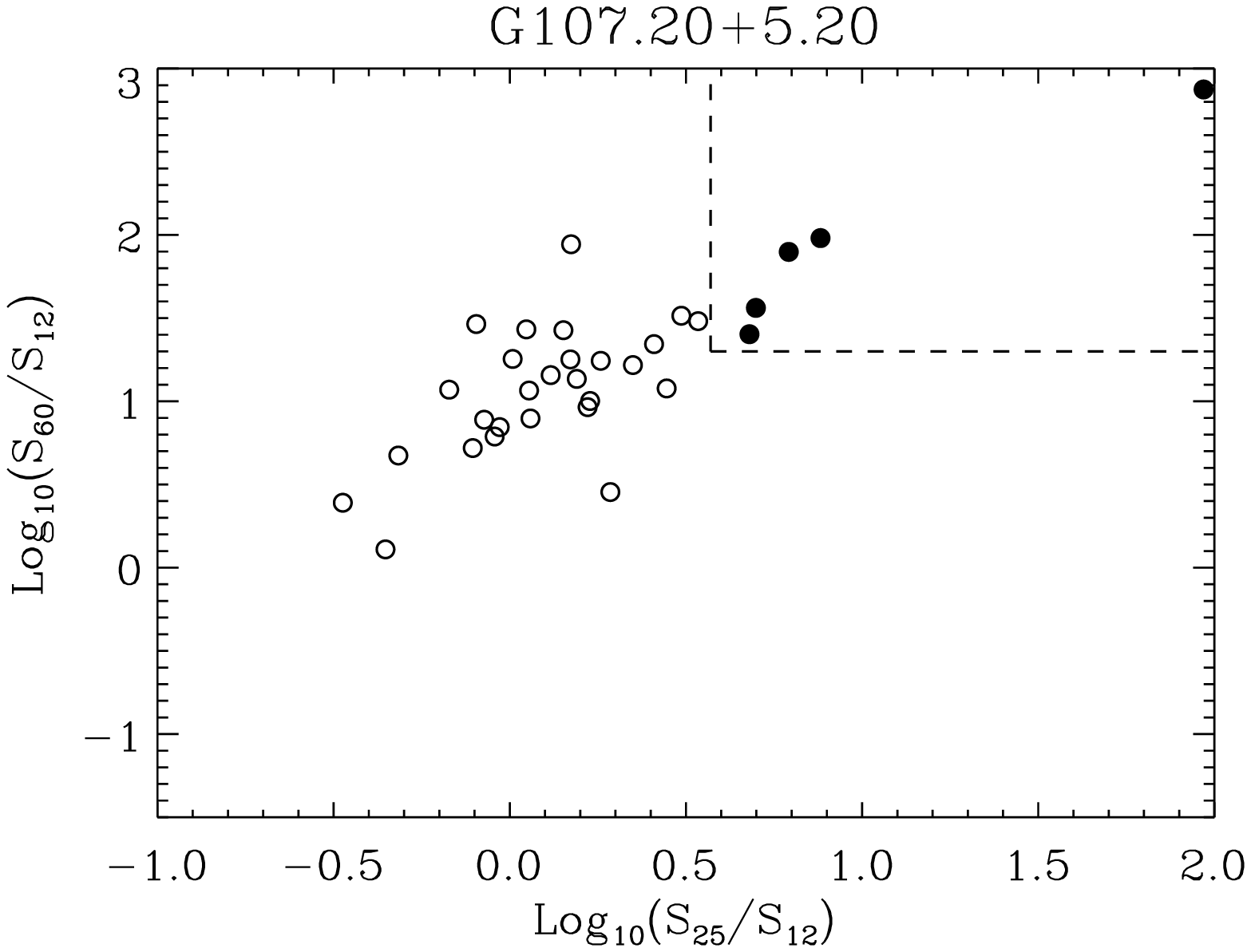}
    \includegraphics[width=5.4cm, angle=0]{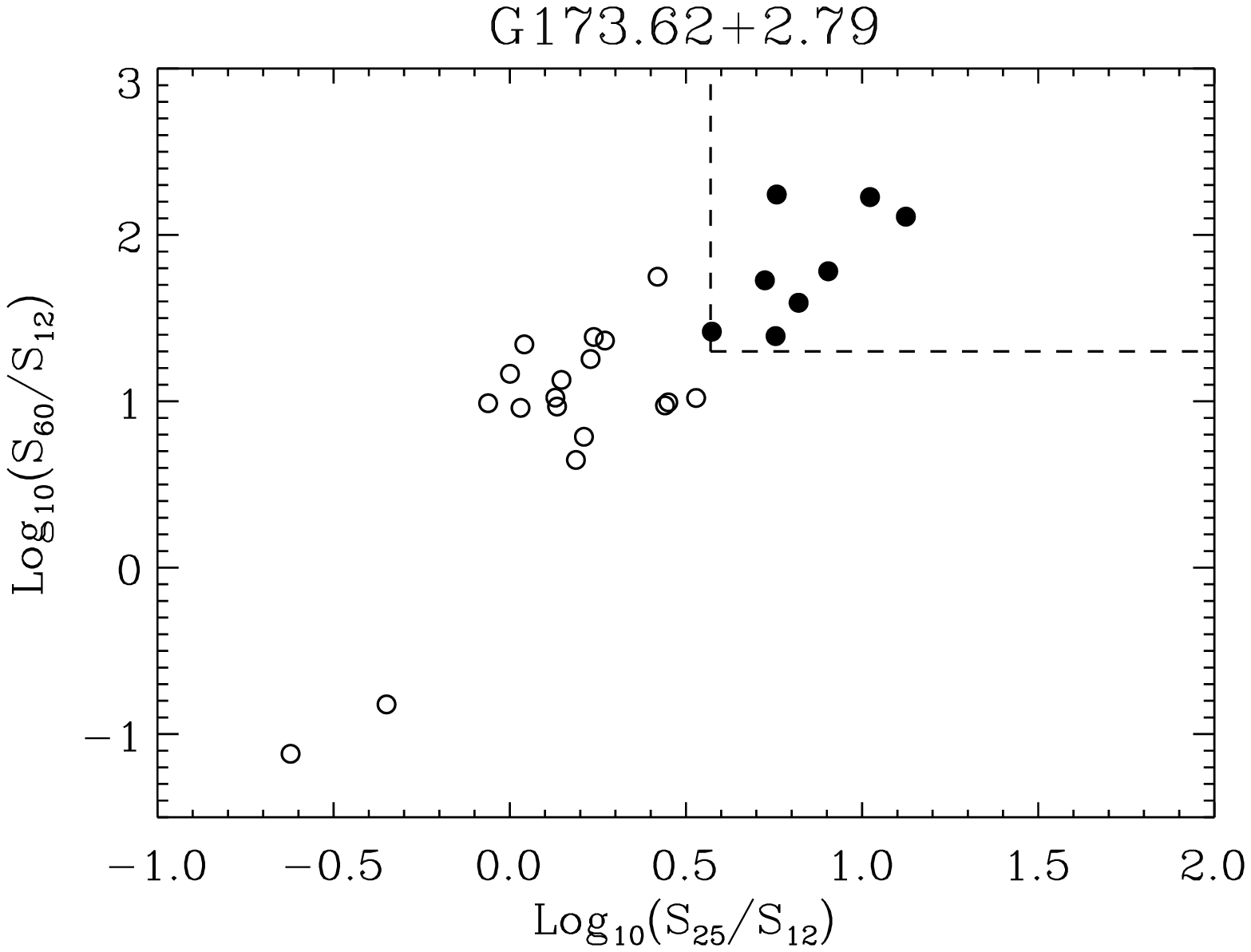}
\caption{Colour-Colour plots of IRAS sources in the vicinity of Perseus ({\it left}), G107.2+5.2 ({\it middle}) and G173.6+2.79 ({\it right}) AME regions. The colours are calculated for sources within $1^{\circ}$ of the central source positions using the IRAS PSC. UCHII region candidates, shown as solid filled circles, have ratios ${\rm log}_{10}(S_{60}/S_{12}) \geq 1.30$ and ${\rm log}_{10}(S_{25}/S_{12})\geq 0.57$, corresponding to the top-right hand corner of this plot (marked with a dashed line). }
\label{fig:uchii}
\end{figure*}

Fig.~\ref{fig:uchii} shows the colours of matched IRAS Point Source Catalogue (PSC v2.1) for the three AME regions. UCHII candidates have ratios ${\rm log}_{10}(S_{60}/S_{12}) \geq 1.30$ and ${\rm log}_{10}(S_{25}/S_{12})\geq 0.57$, corresponding to the top-right hand corner of this plot (marked with a dashed line). We have ignored sources that are categorised as extragalactic (IRAS IDTYPE 1) or only have upper limits at $25$ or $60\,\mu$m. There are a few UCHII candidates within each of the three regions with a wide range of $100\,\mu$m flux densities. Summing these up for each region, and assuming $S_{100\,\mu{\rm m}}/S_{\rm 2\,cm}=1000$ gives 0.52, 14.3 and 7.9\,Jy, for Perseus, S140 and S235, respectively. This corresponds to upper limits of the fraction of the AME that could be due to UCHII at frequencies $\sim 15$--30\,GHz of $<4\,\%$ (Perseus), $<102\,\%$ (S140) and $<122\,\%$. In the Perseus source, the contribution of UCHII is negligible. But for the AME detected in the two bright HII regions (S140 and S235), it could potentially all be due to UCHII. However, this is a very conservative upper limit. High resolution observations (e.g. with AMI at 15\,GHz; see Perrott et al., this issue) shows that the majority of the AME is in fact diffuse and therefore is unlikely to be dominated by UCHII. Nevertheless, the possible contribution from UCHII regions should not be overlooked.

\subsection{AME emissivity of HII regions}

The first detections of dust-correlated AME originate from CMB experiments measuring the sky at high Galactic latitudes, and thus authors have often calculated the dust ``emissivity'' in terms of the radio brightness relative to a dust template map. This has led to the use of the IRAS $100\,\mu$m map being used as a predictor of the AME amplitude with the emissivity defined in these terms; specifically, in units of $\mu$K/(MJy/sr). Typical values for diffuse cirrus at high Galactic latitudes are $\sim 10\,\mu$K/(MJy/sr) with variations of a factor of $\sim 2$ \cite{Davies2006}; this corresponds to approximately 1\,Jy at 33\,GHz for every 3000\,Jy at $100\,\mu$m.

Table~\ref{tab:obs} lists the AME dust emissivities, $E$, in terms of the AME brightness temperature relative to the $100\,\mu$m brightness, converted to units $\mu$K/(MJy/sr). Although the uncertainties are large, it appears that the AME emissivity are comparable to the high latitude value, but on average are lower compared to the high latitude value (and lower still than the Perseus AME region). More strikingly, upper limits from the compact sample observed by AMI \cite{Scaife2008} and also upper limits from W40 suggest the AME emissivities are at least an order of magnitude lower still.

This surprising result can be understood in terms of the different environments in the vicinity of HII regions. The most important is the lack of smallest dust grains (PAHs) which are known to be depleted inside HII regions (e.g. \cite{Povich2007}). If AME is due to spinning dust grains this would severely reduce the AME brightness since the very smallest grains produce most of the spinning dust flux. Other factors may also be contributing such as the interstellar radiation field and distribution of ions. 

Finally, we point out that although the $100\,\mu$m emissivity is a convenient quantity, it can be significantly biased in regions with a higher than average dust temperature. The $100\,\mu$m intensity is very sensitive to the dust temperature. For example, compared to an average dust temperature of 18.1\,K, for a value of 22\,K the $100\,\mu$m intensity is a factor of $4$ times higher, while for 30\,K it is a factor of 23 times higher (see Tibbs et al., this issue). HII regions are known to have warmer dust, typically 30--80\,K, and thus the AME emissivity will naturally be lower. This may explain the apparently lower values observed in more compact regions where this will be pronounced. A better definition of emissivity would be to use the column density or, equivalently, the thermal dust optical depth \cite{Finkbeiner2004}.

\section{Conclusions}
\label{conclusions}

HII regions are an interesting place to look for AME. So far, there have been a number of detections from HII regions (or in the vicinity of HII regions). However, measuring accurate SEDs over a wide frequency range, in addition to the complex environment and the presence of bright continuum (e.g. free-free, thermal dust) emission makes this is a very difficult task. Furthermore, the presence of optically thick free-free emission from UCHII regions may be contributing to a portion of the AME for some regions. Nevertheless, the AME emissivity is comparable to the more robust detections on molecular clouds and diffuse cirrus, although on average it is lower; in some HII regions there are only upper limits on AME. More data, particularly at higher resolutions and high frequencies ($>5$\,GHz) are needed.

\vspace{1cm}
{\it Acknowledgements.} CD acknowledges an STFC Advanced Fellowship, an EC Marie-Curie IRG grant under the FP7, and an ERC Starting Grant (No. 307209).

\bibliographystyle{abbrv}
\bibliography{refs}

\end{document}